\documentclass[12pt]{article}
\usepackage{graphicx}
\usepackage{color}
\usepackage[dvipsnames]{xcolor}
\usepackage[normalem]{ulem}
\usepackage{amsmath,slashed, amssymb}
\usepackage{fancyhdr}
\usepackage[linktocpage]{hyperref}
\usepackage[normalem]{ulem}
\usepackage[titletoc]{appendix}
\numberwithin{equation}{section}

\def\beq{\begin{eqnarray}}
\def\eeq{\end{eqnarray}}
\def\bea{\begin{eqnarray*}}
\def\eea{\end{eqnarray*}}

\def\centeron#1#2{{\setbox0=\hbox{#1}\setbox1=\hbox{#2}\ifdim
\wd1\rangle\wd0\kern.5\wd1\kern-.5\wd0\fi
\copy0\kern-.5\wd0\kern-.5\wd1\copy1\ifdim\wd0\rangle\wd1
\kern.5\wd0\kern-.5\wd1\fi}}
\def\ltap{\;\centeron{\raise.35ex\hbox{$\langle$}}{\lower.65ex\hbox{$\sim$}}\;}
\def\gtap{\;\centeron{\raise.35ex\hbox{$\rangle$}}{\lower.65ex\hbox{$\sim$}}\;}

\newcommand{\newc}{\newcommand}
\newc{\qbar}{{\overline q}}
\newc{\Kahler}{Kahler }
\newc{\deltaGS}{\delta_{\rm GS}}

\begin{document}
\begin{titlepage}
\begin{flushright}
{\large SCIPP 20/08
}
\end{flushright}

\vskip 1.2cm

\begin{center}

{\LARGE\bf Obstacles to Constructing de Sitter Space in String Theory}

\vskip 1.4cm

{\large Michael Dine{\it $^a$}, Jamie A.P. Law-Smith{\it $^b$}, Shijun Sun{\it $^a$}, Duncan Wood{\it $^a$}, and Yan Yu{\it $^a$}}
\\
\vskip 0.4cm
{\it $^{(a)}$Santa Cruz Institute for Particle Physics and \\
Department of Physics, University of California at Santa Cruz \\
1156 High St, Santa Cruz, CA 95064, USA} \\
~\\
{\it $^{(b)}$Department of Astronomy and Astrophysics, University of California at Santa Cruz \\
1156 High St, Santa Cruz, CA 95064, USA} \\
\vspace{0.3cm}

\end{center}

\vskip 4pt


\begin{abstract}
There have been many attempts to construct de Sitter space-times in string theory. While arguably there have been some successes, this has proven challenging, leading to the de Sitter swampland conjecture: quantum theories of gravity do not admit stable or metastable de Sitter space. Here we explain that, within controlled approximations, one lacks the tools to construct de Sitter space in string theory. Such approximations would require the existence of a set of (arbitrarily) small parameters, subject to severe constraints. But beyond this one also needs an understanding of big-bang and big-crunch singularities that is not currently accessible to standard approximations in string theory. The existence or non-existence of metastable de Sitter space in string theory remains a matter of conjecture.
\end{abstract}

\end{titlepage}

\tableofcontents

\section{Introduction: The de Sitter Swampland Conjecture}

The observable universe appears to have emerged from a period of high
curvature.  Almost certainly, if we run the clock backwards, we encounter a period
where classical general relativity does not apply.  Remarkably, while string theory has provided tools to think about many questions in quantum gravity, cosmologies resembling
our own remain inaccessible to controlled approximations in the theory.  Conceivably the observed big bang is not described by a quantum theory of gravity or requires some still larger structure, but it would seem more likely that this simply represents a failure of our present collection of theoretical tools.

Strong evidence from supernovae \cite{Perlmutter:1998np}, CMB \cite{2018arXiv180706209P}, and Large Scale Structure observations \cite{2010MNRAS.401.2148P} suggest that our universe has entered a stage of exponential expansion, well-described as a de Sitter solution of Einstein's equations.  At a time shortly after the big bang, there is good reason to think that the universe also went through a period of exponential expansion \cite{Guth:1980zm, Starobinsky:1980te, LINDE1982389, Albrecht:1982wi}.  So de Sitter space seems likely to play an important role in any understanding of our present and past universe.  The inflationary period lasted only for a brief moment; our limited understanding of how de Sitter space might
arise in string
theory would suggest that even our present de Sitter universe is metastable.  

The notion of a {\it cosmic landscape} introduces another role for spaces of positive cosmological
constant (c.c.).  In particular, such a landscape might allow a realization of anthropic selection
of the c.c. \cite{weinbergcc}, but would seem to require the existence of a vast set of metastable, positive c.c. vacua.

Given these considerations, the conjecture of \cite{ovswampland} that metastable de Sitter space lies in the swampland of quantum gravity theories is particularly interesting, with possible implications for inflation, the nature of the currently observed dark energy, and
implementing the anthropic explanation of the c.c.  We will not address the conjecture in its full generality, but we will examine the starting point.  The authors of \cite{ovswampland} begin with the observation that it has proven difficult to construct de Sitter space in string theory.  While there are constructions that appear to achieve a positive cosmological  stationary  point in a suitable effective action \cite{andriot,andriot2019}, it is not clear that they are in any sense generic.

But one should first ask:  what would it mean to construct de Sitter space in string theory?  In most constructions, one starts with some classical solution of the equations of critical string theory. 
These solutions invariably have moduli or pseudomoduli.  Then one adds features, such as fluxes, branes, and
orientifold planes which give rise to a potential for these moduli, and looks for a local minimum with positive four-dimensional c.c.  These attempts to construct de Sitter space generally raise two questions.  First, what is the approximation
scheme that might justify any such construction?  Second, any would-be de Sitter space
found in this way is necessarily, at best, metastable:  inevitably there is a lower energy density in asymptotic regions of the original moduli space.  Quantum mechanically, the purported de Sitter state cannot be eternal.  It has a history; it will decay in the future and must have been created by some mechanism in the past.  The quantum mechanics of this process is challenging to pin down.  In this paper, we will see that already classically, the notion of an eternal de Sitter space in string theory is problematic; small perturbations near the
de Sitter stationary point of the effective action evolve to singular cosmologies.

In more detail, there are at least two challenges to any search for metastable de Sitter space in string theory:
\begin{enumerate}
\item  One requires a small parameter(s) allowing a controlled approximation to finding stationary points of an effective action.  Here one runs into the problem described in \cite{dineseibergproblem}.  Without introducing additional, fixed parameters (i.e., introducing parameters not determined
by moduli), would-be stationary points in the potential for the moduli lie at strong coupling.  Typically, attacks on this problem (and the question of de Sitter space) exploit large fluxes\footnote{The KKLT \cite{kklt} constructions are, in some sense, an exception, which we will discuss later.}.  If there is to be a systematic approximation, it is necessary that the string coupling be small and compactification radii large at any would-be stationary point found in this way.  If the strategy is to
obtain inverse couplings and radii scaled by some power of fluxes, it is also important that these fluxes (and possibly other discrete parameters) can be taken arbitrarily large, without spoiling the effective action treatment.  Even allowing uncritically for this latter possibility, we will see that it is quite challenging to realize arbitrarily weak string coupling and large radius, with positive or {\it negative} c.c.\footnote{This point has been noted
earlier \cite{junghans1, junghans2,wrase}.  A broad critique, applicable
to many non-perturbative scenarios,
has been put forward in \cite{sethi}.} 
\item  If one finds such a stationary point, one must ask about stability.  More precisely, in string theory, we are used to searching for suitable background
geometries and field configurations 
by requiring that the evolution of excitations about these configurations is described by a unitary
$S$ matrix.  Classically, at least in a flat background, this is the statement that any initial perturbation of the system has a sensible evolution to some final perturbation.  Again, we will see that this requirement is problematic for any would-be classical de Sitter stationary point in such
a theory; even if all eigenvalues of the mass-squared matrix (small fluctuation
operator) are 
positive,  large classes of small perturbations evolve to singular geometries. 
\end{enumerate}

The problem of evolution of small perturbations is connected with the properties of the moduli of string compactifications, described above.  We consider, in particular, disturbances of the moduli fields in a classical, eternal de Sitter space.   We will see in this paper that some small fluctuations in the far past are amplified, rolling over the barrier to a contracting universe that culminates in a big crunch singularity.  As a result, already classically, there is no notion of an $S$ matrix (in the sense of describing the future of any small disturbance of the system), even restricted to very small perturbations localized near the metastable minimum of the potential.  Within our current collection of calculational tools, we lack any framework in string theory to study such singularities.  As a result, we will explain, the problem of constructing de Sitter space in string theory is not, at least at present, accessible to systematic analysis.

Overall, then, we will argue that we lack theoretical methods to address, in any systematic fashion, the problem of constructing de Sitter space in string theory, much as we lack the tools to understand
big bang or big crunch singularities in any controlled approximation.   The existence of metastable de Sitter states may be plausible or not, but it is a matter of speculation.\footnote{Reference \cite{valenzuela} gives non-perturbative arguments for the absence of de Sitter vacua in controlled approximations. Various scenarios for how
de Sitter might arise, and how this might be understood, even lacking a systematic approximation, have been put forward.  Among many examples, \cite{garg1,garg2} argue for a more refined version, based on explicit constructions;
\cite{heckman1,heckman2} consider F-theory compactifications and associated prospects.  \cite{geng} proposes another way in which de Sitter might
arise.   \cite{march-russell} takes a phenomenological view of the problem.
An alternative discussion of de Sitter space in flux vacua appears in \cite{sethi}, who argues against flux stabilization on rather general
grounds.  \cite{cicoli1} takes an optimistic view of the prospects
for such constructions and \cite{cicoli2,cicoli3,cicoli4} put
forth several scenarios.}    The failure to find such states in any controlled analysis appears, at least at present, inevitable.

\section{The $S$ Matrix and Classical Field Evolution}

Much of our focus will be on the evolution of classical perturbations in metastable de Sitter space.
We will argue that many of these perturbations evolve towards a big crunch singularity, and that this is outside
of the scope of current methods in string theory/quantum gravity. 
In critical string theory, the object of interest is the $S$ matrix.  A {\it classical} solution of the string equations corresponds to a space-time for which one can define a sensible scattering matrix.  The connection to classical scattering, in field theory and string theory, arises from considering the evolution of small disturbances.  These correspond to initial {\it and final} isolated, localized states, with large occupation numbers.  These can be considered as coherent states.  For a single real scalar field, for example, one can develop a classical perturbation theory.   Start, 
at lowest order, with a field configuration of the form
\beq
\phi(x) = \phi_{\vec p_1}(x) + \phi_{\vec p_2}(x)+
 \phi_{\vec k_1}(x) + \phi_{\vec k_2}(x)
\eeq
where each term represents a localized wave packet with mean momentum $\vec k_i$. 
Momentum conservation requires $\vec p_1 + \vec p_2 = \vec k_1 + \vec k_2$
within the momentum uncertainty,  and non-trivial scattering requires that the wave packets all overlap
at a point in space-time.  Quantum mechanically, the scattering problem we have outlined here corresponds to some large number of particles of each momentum in both the initial
and final states.
Making a decomposition into positive and negative frequency components:
\beq
\phi(x) = \phi^+(x) + \phi^-(x) \rightarrow \begin{cases} ~\phi^+(x) \vert \Phi \rangle = \Phi(x)\vert \Phi
\rangle \\ ~\langle \Phi \vert \phi^-(x) =\langle \Phi \vert \Phi^*(x)\end{cases} .
\eeq
In momentum space, $\Phi^\pm(\vec k)  e^{i \vec k \cdot \vec x \mp i \omega t}$ corresponds to the positive and negative frequency components.  Reality requires $\Phi^\pm(\vec k) = \Phi^{\pm *}(-\vec k).$  Occupation numbers scale as $\vert \Phi^\pm(\vec k)\vert $.

Order by order in the interaction, $\lambda \phi^4$, we can compute corrections to the classical
scattering, 
\beq
\delta \phi(x) = \delta \phi_{\vec p_1}(x) + \delta \phi_{\vec p_2}(x)+
 \delta \phi_{\vec k_1}(x) +\delta \phi_{\vec k_2}(x).
\eeq
Evaluated at the interaction point, $\delta \phi$ defines an $S$ matrix (more precisely a $T$ matrix) on the space of coherent states.  This can be decomposed as an $S$ matrix on states of definite particle number; the classical approximation is valid when the occupation numbers are large.

Phrased this way, the statement that one can construct an $S$ matrix for large occupation numbers in
initial and final states is the statement that one has sensible evolution from any initial classical configuration (described by $\vec p_1,\vec p_2$) to any final configuration ($\vec k_1,\vec k_2$).  

In the case of de Sitter space, the question of the existence of an $S$ matrix is subtle \cite{srednickietal}.  We will focus, instead, on what we view as a minimal requirement that all classical perturbations in a would-be metastable de Sitter vacuum have a sensible evolution
arbitrarily far into the future.  We will
see that some subset of possible perturbations evolve to singular geometries, over which we have  no
theoretical control.  We argue that this means that one does not have a controlled construction
of such spaces.   The existence, or not, of such metastable de Sitter spaces then becomes a
matter of conjecture.

\section{Searching for Stationary Points of an Effective Action}

We first explore some of the challenges
to the construction of stationary points of the effective action
with positive c.c.  Typically, these efforts involve the introduction of branes, orientifold planes, and fluxes \cite{andriot}.  One searches
for particular stationary points of the action with positive
cosmological constant, and asks whether
the string coupling is small and the compactification radii
large at these points \cite{andriotwrase1,andriotwrase2}.  This, by itself, does not address the question
of whether there is a systematic approximation.  The system with branes
and fluxes is not a small perturbation of the system without,
and the range of validity of the expansion in one is not related
to that of the other.  If there is to be a systematic
approximation of any sort, one requires a sequence of such stationary
points as one increases the flux numbers; the would-be small parameters are the inverse of some large flux numbers.
In our discussion we will assume that it makes sense to take such numbers arbitrarily large.  Then the goal
is to find stable, stationary points of the action where
\begin{enumerate}
\item  The string coupling is small.
\item  All compactification radii are large.
\item  The cosmological constant is small and positive.
\end{enumerate} 
As reviewed in \cite{andriot2019}, satisfying this set of constraints is challenging. We review some of the issues in this section.  Similar
analyses, with similar conclusions, have appeared in \cite{junghans1,junghans2,wrase}.  Our point of view is that this is not surprising.  Searches at weak coupling were not likely to
yield non-supersymmetric
metastable vacua, dS or AdS,
and provide little information about the
existence or non-existence of such states.  For the dS case, it is hard
to see how such states could be understood without a much broader understanding of their cosmology, as we will discuss subsequently.

We follow \cite{andriot} in studying type II theories in the presence of an $O_p$ plane,
and a background geometry with metric
\beq
ds^2 = g_{\mu \nu} dx^\mu dx^\nu +\rho~ g_{IJ}^0 ~ dy^I dy^J .
\eeq 
Here $g_{IJ}^0$ represents a background reference metric for the compactified dimensions.
$g_{\mu \nu}$ represents the metric of four dimensional space-time, which we hope to be de Sitter.
Reference \cite{andriot} distinguishes directions parallel and perpendicular to the orientifold plane with an additional modulus $\sigma$; for simplicity, we assume $\sigma \sim 1$; this assumption can be relaxed without severe difficulty. We ignore other light moduli as well. 
We also include NS-NS $3$-form and R-R $q$-form fluxes, $H_{IJK}^{(n)}, ~F_q^{(n)}$.  

The fluxes will be understood as taking discrete, quantized values.  The dependence of terms  
on the moduli $\rho$ and $\tau = \rho^{3/2}~ e^{-\phi}$ is given in \cite{andriot}, and is
readily understood from the following considerations:
\begin{enumerate}
\item  In the NS-NS sector, there is a factor $1/g^2 = e^{-2\phi}$ in front of the
action.  The four-dimensional Einstein term has a coefficient $\tau^2 = \rho^{3} / g^2$.  This can be brought to canonical form by
the Weyl rescaling, $g_{\mu \nu} \rightarrow g_{\mu \nu} \tau^{-2}$.  The moduli
$\tau$ and $\rho$ will be our focus.
\item  Again in the NS-NS sector, terms involving the three-index tensor, before rescaling, contain a factor $\tau \rho^{-3}$; after the Weyl rescaling, they acquire an additional factor of $\tau^{-3}$ in front.  Terms involving the
six-dimensional curvature similarly scale as $\rho^{-1} \tau^{-3}.$
\item  In the RR sector, the flux terms have, before rescaling, no  factors of $1/g$.  They have various factors of $\rho$ depending on the rank of the tensor.  The Weyl rescaling introduces
a factor of $\tau^{-4}$.
\end{enumerate}
The resulting action is \cite{andriot}:
\begin{equation}
\begin{split}
V &= -\tau^{-2}\left(\rho^{-1}R_6(\sigma) - \frac{1}{2}\rho^{-3}\sum_{n}\sigma^{6n-3(p-3)}\left|H^{(n)}\right|^2\right)-\tau^{-3}\rho^{\frac{p-6}{2}}\sigma^{\frac{(p-3)(p-9)}{2}}\frac{T_{10}}{p+1}\\ & + \frac{1}{2}\left(\tau^{-4}\sum_{q=0}^4\rho^{3-q}\sum_{n}\sigma^{6n-q(p-3)}\left|F_q^{(n)}\right|^2 +\frac{1}{2}\tau^{-4}\rho^{-2}\sum_{n}\sigma^{6n-5(p-3)}\left|F_5^{(n)}\right|^2\right) .
\end{split}
\end{equation}
Again, we will ignore the index $(n)$ in what follows and set $\sigma =1$.
To illustrate the issues, we will consider large $F_2$ and $F_4$.   These fluxes
satisfy,
with $H_3=0$, Bianchi identities, with a source for $F_4$. These
equations can be satisfied with large fluxes through two and four
cycles.  

For
$3 \leq p \leq 7$ 
and choosing $T_{10} =1$, $R_6\sim 1$, we can drop the $T_{10}$ term because the $R_6$ term will dominate. We can attempt to find large $\tau$ and $\rho$ by turning on $F_2 = n_2$ and $F_4= n_4$ (other combinations of fluxes give similar results). 
Then one has the relevant terms:
\beq
-\tau^{-2} \rho^{-1} R_6+ \frac{1}{2}\tau^{-4} \left(n_2^2 \rho + n_4^2 \rho^{-1} \right).
\eeq
Differentiating with respect to $\rho$ and $\tau$,  for $n_4 \gg n_2 \gg 1$, one has then 
\beq
\rho^{-2} R_6 + \frac{1}{2}\tau^{-2} \left(n_2^2 - n_4^2 \rho^{-2} \right) = 0
\eeq
and
\beq
\rho^{-1} R_6 - \tau^{-2} \left(n_2^2 \rho + n_4^2 \rho^{-1} \right) = 0.
\eeq
We get a solution of the form:
\beq
\rho^2  =-\frac{1}{3}\left ( {n_4 \over n_2} \right )^2; ~\tau^2 = \frac{2}{3} {n_4^2 \over R_6}.
\eeq
Negative $\rho^2$ is not acceptable.  But even if somehow $\rho^2$ had been positive, we would have had:
\beq
g^2 = {\rho^{3} \over \tau^2}  
\propto R_6 \left ( {n_4 \over   n_2^3} \right );
\eeq
so the string coupling would not have been weak.
The other terms we have neglected are suppressed at this point.  For example, the term
proportional to $T_{10} \rho^{-3/2} \tau^{-3}$ is suppressed by $(n_2 / n_4)^{2}$.

For $p=8$, which corresponds to the $T_{10}$ term dominating, turning on, again, $n_4$ and $n_2$, one finds that $\rho^2 = -7 n_4^2 / n_2^2$, which is also negative. Parameterically, one now has $g^2 \propto n_4^3 / n_2^7$, so again, even if one ignored signs, this regime would give large $\rho$ and $\tau$ but also large $g$.

An interesting case is provided by $p=8$ with $n_0$ and $n_2$ non-zero.  In this case, one finds that
\beq
\rho^2 = {1 \over 5} {n_2^2 \over n_0^2}~; ~~\tau = {8 \over 5} n_2^2
\eeq
so one requires $n_2 \gg n_0$.  Both quantities are now positive, but the cosmological constant,
consistent with expectations of \cite{andriot}, is negative, corresponding to AdS space.  Setting this aside, one has that
\beq
g_s^2 \propto {1  \over n_2^2}
\eeq
so the string coupling is small.  But this is not good enough.  If one considers higher derivative
terms in the effective action at tree level ($\alpha^\prime$ expansion) these are not suppressed.
Writing the action in ten dimensions, the terms (written schematically)
\beq
\int d^4 x d^6 y \sqrt{g_4} \sqrt{g_6} \left( F_{IJ} F^{IJ} + \left( F_{IJ} F^{IJ} \right)^2 \right)
\eeq
are both of the same order in the large flux, $n_2^2$, due to the two extra factors of $\rho^{-2}$
coming from the two extra powers of inverse metric in the second term.
For all values of $p$, if we just consider the $H$ and $F_q$ terms, $\partial V/\partial \tau =0$ gives negative $\rho^2$. 

In other cases, one finds these and other pathologies---AdS rather than dS stationary points and instabilities.
Searches involving broader sets of moduli \cite{andriotwrase1,andriotwrase2} seem to allow at best a few isolated regions of parameter space where such solutions might exist.  Whether these might exhibit
a sensible perturbation expansion is currently an open question, but our results
above suggest that the combination is a tall order.  So, even with the large freedom in flux choices we have granted ourselves,
metastable de Sitter stationary points would appear far from generic in regimes where
couplings are small and compactification radii are large.

\section{Expectations for Evolution of Perturbations in de Sitter Space}

String theory has had many dramatic successes in understanding issues in quantum gravity.  But one
severe limitation is its inability, to date, to describe cosmologies resembling our own, which {\it appear} to emerge from a big bang singularity or evolve to a big crunch singularity.  
This
could reflect some fundamental limitation; more likely, it reflects the inadequacy of our present
theoretical tools to deal with situations of high curvature and strong coupling.
For example, consider a pseudomoduli space where the potential falls to zero for large fields
in the positive direction. If one starts the system in the far past with expanding boundary conditions,
then further in the past there is a big bang singularity; if one starts with contracting boundary
conditions, there is a big crunch in the future \cite{banksdinemodulievolution}. These high curvature/strong coupling regions are inevitable, despite the system being seemingly weakly coupled through much of this history.  It is possible that in any string 
cosmology, there need not be an actual curvature singularity, but the growth 
of the curvature means that the system enters a regime where any conventional sort of effective action or conventional weak coupling string description breaks down. 
It seems hard to avoid the 
conclusion that there is such a singularity (regime of high curvature) in the past or future of cosmological solutions on 
a moduli space.  These problems might be avoided in some more complete treatment of the problem within the framework of a single cosmology, or perhaps something else, such as eternal inflation in a multiverse, is needed. In any case, the problem is beyond our present theoretical reach.

Our question, in this section, is: {\it are things better for metastable de Sitter space-times}?  In particular, in efforts to construct de Sitter space-times in string theory, the strategy is to search some effective action for a positive c.c. stationary point, separated by a finite potential
barrier from
a region in field space where, asymptotically, the potential tends to zero.  If we start the system
at the local minimum of the potential, classically, it will stay there eternally.  But how do small
fluctuations evolve?  Might there be small disturbances that 
drive the field to explore the region on the other side of the barrier,
exhibiting the pathologies of the system on pseudomoduli spaces of \cite{banksdinemodulievolution}?

In one presentation of de Sitter space (which covers all of the space):
\beq
ds^2 = -d\tau^2 + \cosh^2(H\tau) \left[ d\chi^2 + \sin^2 \chi d\Omega_2^2 \right].
\label{kequalsoneds}
\eeq
A homogeneous scalar field in this space, $\phi(\tau)$, obeys
\beq
\ddot \phi + 3 H{\sinh (H\tau) \over \cosh(H\tau)} \dot \phi + V^\prime(\phi) = 0.
\eeq
The equation is slightly more complicated if $\phi$ depends on $r$ as well.

The metric of equation \ref{kequalsoneds} respects an $SO(4,1)$ symmetry, as well as a $Z_2$ that reverses the sign of $\tau$.
Suppose, first, the potential for $\phi$ rises in all directions about a minimum (taken at $\phi=0$ for simplicity). For large positive $\tau$, any perturbation of $\phi$ about
a local minimum damps; for large negative $\tau$, the motion is amplified as $\tau$ increases (it damps out in the past).  Correspondingly, in the far past and the far future, the field approaches the local minimum (to permit a perturbative discussion, we must require that the maximum
value of the disturbance at all times is small).  Starting in the far past, we can think in terms of
a localized disturbance in space (e.g., due to a source localized in space time) and study
the Fourier transformed field.  If the disturbance has some characteristic momentum $k$, this
momentum will blueshift exponentially as $\tau \rightarrow 0$, and the
amplitude will grow.  For $\tau>0$, the distribution will damp and redshift to longer wavelengths.  

If the perturbation
has scale smaller than $H^{-1}$ (and in particular if the Hubble constant is small compared to the curvature
of the potential), then the space-time near the disturbance is approximately flat, and, assuming rotational invariance, the disturbance 
breaks $SO(3,1) \times {\rm translations}$ to $SO(3)$. In terms of the full symmetry of de Sitter space, the
perturbation breaks $SO(4,1)$ to $SO(3)$. 
To summarize, any approximately
homogeneous disturbance in eternal de Sitter corresponds to a solution that grows in the far past and decreases
in the future.  One can define {\it past} and {\it future} relative to the point where the scalar field is a maximum.  The location of this point breaks much of the continuous symmetry of de Sitter space
but leaves $SO(3) \times Z_2$, where the $Z_2$ represents time reversal about the point
where the amplitude of the field oscillation is a maximum.  The maximum of the field, indeed, provides a natural definition
of the origin of time.  At this point, the time derivative of the field vanishes. 

Now for a potential that has a local minimum with positive energy density, and that falls to zero for large $\vert \phi \vert$, we might expect that if we create a small, localized perturbation at some ($r_0, \tau_0$)
this perturbation will damp out if $\tau_0 \gg 0$.  But if $\tau_0 \ll 0$, the perturbation will grow, possibly crossing over the barrier while $\tau \ll 0$.  In this case, the emergent universe on the other side of the barrier is contracting, and
we might expect the system to run off towards $\phi = \infty$, until the universe undergoes
gravitational collapse.  If this is the case, then the $Z_2$ symmetry might be said to be spontaneously broken; one has a pair of classical solutions, one with a singularity in the past, one in the future, related by the $Z_2$ symmetry.

Before establishing this fact, it is helpful to review some aspects of the Coleman-De Luccia (CDL) bounce from this perspective \cite{cdl}.

\subsection{The Coleman-De Luccia bounce as a solution of the field equations with Minkowski signature}

We are interested in disturbances which lead to motion over a barrier, rather than
tunneling.  We might expect, however, that once the system passes over the barrier,
its subsequent evolution is not particularly sensitive to whether it passed over the barrier 
or tunneled through it. In the case of a thin-wall bubble, before including gravity, at large times,  the bubble wall becomes relativistic, and the
bubble radius is of order $t$, so  one expects that the bubble
energy is proportional to $t^3$, dwarfing any difference in the energy of order the barrier height at the
time of bubble formation.  The same is true for a thick-walled bounce connecting
two local minima of some potential.  
In other words, at large time, at least for very small $G_N$,  we might expect the solution to be a small
perturbation of the bounce solution of Coleman \cite{colemandecay}
and Coleman and De Luccia \cite{cdl}, which we will review briefly.

\subsection{Tunneling with $G_N = 0$}

Consider, first, the bounce solution without gravity.  We consider a potential, $V(\phi)$, with local minima at $\phi_{\rm true},~\phi_{\rm false}$, where $V(\phi_{\rm false}) > V(\phi_{\rm true}).$  Starting with the field equations,
\beq
\Box \Phi + V^\prime(\phi) =0,
\eeq
for points that are space-like separated from the origin (the center of the bubble at the
moment of its appearance),  we introduce $\xi^2 = r^2 -t^2$, in terms of which
\beq
{d^2 \phi \over d\xi^2} + {3 \over \xi}{d\phi \over d\xi} -V^\prime(\phi) =0. 
\eeq
This is the Euclidean equation for the bounce.

For points that are time-like separated, calling $\tau^2 = t^2 - r^2$,
\beq
{d^2 \phi \over d\tau^2} + {3 \over \tau}{d\phi \over d\tau} +V^\prime(\phi) =0. 
\eeq
These equations are related by $\xi = i \tau$.  

On the light cone, $\xi = \tau =0$, we have $d\phi / d\tau = d\phi / d\xi =0$, and we have
to match $\phi(0) = \phi_0$.  In the tunneling problem \cite{colemandecay}, $\phi_0$ is determined by the requirement that $\phi \rightarrow \phi_{\rm false}$ as $\xi \rightarrow \infty$; this can be thought of as a requirement of finite energy relative to the configuration where $\phi=\phi_{\rm false}$ everywhere.  

Independent of the quantum mechanical tunneling problem, the bounce is a solution of the source-free field equations for all time (positive and negative) and everywhere in space.  In the time-like region, the solution
for negative time is identical to that for positive time.  Translation invariance is broken, but
$SO(3,1)$ invariance and the $Z_2$ invariance are preserved.

\subsection{Classical perturbations of the false vacuum with $G_N=0$}

Without gravity, we might consider starting the system in the false vacuum and giving it a ``kick'' so that, in a localized region, the system
passes over the barrier.  On the other side, the system looks like a bubble, but not of the critical size.
We might expect that the evolution of the bubble, on macroscopic timescales, is not
sensitive to the detailed, microscopic initial conditions.  For a thin-walled bubble, for example,
we can think of configurations, as in \cite{colemandecay}, where at time $t=0$, one has a bubble
of radius $R_0$, inside of which one has true vacuum, outside false vacuum, and a transition region described by the kink solution of the one dimensional field theory problem with nearly degenerate minima.
Take the case of a single real field, $\phi$, with potential:
$$
V (\phi) = -{1 \over 2} \mu^2 \phi^2 + {1 \over 4} \lambda \phi^4 + \epsilon \phi + V_0.
$$
For small $\epsilon$, the minima of the potential lie at
\beq
\phi_\pm \approx \pm \sqrt{\mu^2 \over \lambda} .
\eeq 
We can define our bubble configuration, with radius $R$ large compared $\mu^{-1}$, as the kink solution of the one dimensional
problem,
\beq
\phi_B(r;R)= {\phi_+ - \phi_- \over 2} \tanh\left({\mu (r-R) \over \sqrt{2}} \right) + {\phi_+ + \phi_- \over 2} .
\eeq
For our problem, we want to treat $R \rightarrow R(t)$ as a dynamical variable.   If
$R_0(t)$ is slowly varying in time (compared to $\mu^{-1}$), then we can write an action for $R$, 
\beq
S = \int dt \int r^2 dr d\Omega \left({1 \over 2} (\partial_t \phi_B(r;R(t)))^2 -(\vec \nabla \phi_B(r,R(t)))^2 - V(\phi_B(r,R(t)))\right)
\eeq
$$
~~~~\approx   \int dt 4 \pi R^2\int_{R - \delta}^{R + \delta}  dr   \left({1 \over 2}(\partial_{r} \phi)^2\right ) \left (   \dot R^2
- 2 \right ), 
$$
where we have used the thinness of the wall to reduce the three-dimensional integral to a one-dimensional integral,
and the fact that for the kink solution, the kinetic and potential terms are equal, to write the second term.   We will
restore the $\epsilon$ term in a moment.

The integral over the bounce solution is straightforward, yielding $\sqrt{2/3}$. 
So we have the effective action for $R$,
\beq
S = 4 \pi \int dt \left ( \sqrt{2 \over 3} \mu^3(R^2 \dot R^2 - 2 R^2) + {\epsilon \over 3} R^3 \right ).
\eeq
Correspondingly, the energy of the configuration is:
\beq
E(R,\dot R) = 4 \pi \left ( \sqrt{2 \over 3} (R^2 \dot R^2 + 2 R^2 -{1 \over 3} \epsilon R^3) \right ) \equiv {M(R) \over 2} \dot R^2  + V(R).
\eeq
We can extract several results from this expression.  In particular we have:
\begin{enumerate}
\item The point where the potential vanishes, $R = R_1 = {2\sqrt{2}\mu^3 \over \epsilon}$.
\item  The location and value of the potential at the maximum: $R= R_2 = \frac{4 \sqrt{2} \mu^3}{3 \epsilon}$.
\item  We can determine $\dot R$ as a function of $R$ and the initial value of $R$ (for simplicity assuming $\dot R(0) = 0$).
\end{enumerate}

We have checked, numerically, that starting with a field configuration corresponding to $\phi(x,t=0) = \phi_B(r;R), ~\dot \phi(x,t=0) =0$, to the left of the barrier, the bubble collapses.   Starting slightly to the right, the wall quickly becomes
relativistic and expands.  This is consistent with an intuition that the energy of conversion of false vacuum to true is largely converted
into the energy of the wall.

We can make this latter statement more precise.  
If we write:
\beq
\phi(r,t) = \phi_{\rm cr}(t,r) + \chi(t,r), ~~~~~\vert \chi \vert \ll \phi_{\rm cr},
\eeq
where $\phi_{\rm cr}$ is the critical bubble solution,
then
\beq
(\partial^2 + m^2(r,t)) \chi = 0.
\label{chieom}
\eeq
Here $m^2$ is essentially a $\theta$ function, transitioning between the
mass-squared of $\chi$ in the false and true vacua.  Since the bubble
wall moves at essentially the speed of light, and undergoes a length contraction by $t \sim \gamma$, we have that
\beq
m^2(t,r) \approx m^2 (t^2 - r^2)
\eeq
and the $\chi$ equation is solved by
\beq 
\chi = {1 \over r} \chi(t^2 - r^2).
\label{solchi}
\eeq 
So the amplitude of $\chi$ decreases with time, and the energy stored is small compared to that in the bubble wall.

We expect the same to hold for a thick-walled bounce.

\subsection{Behavior of the disturbance with small $G_N$}

Consider the same system, now with a small $G_N$.  Again, our disturbance, after a short period of time, approaches the critical ($G_N=0$) bubble.
At larger time, it will then agree with the Coleman-De Luccia solution, including the small effects of gravity.

As we will see in the next section, for the asymptotically falling potential, with expanding boundary conditions, the evolution of the configuration
is non-singular.  But with contracting boundary conditions, one encounters, as expected, a curvature singularity.

\section{Behavior of the Bounce with Asymptotically Falling Potential}

We have argued that, independent of the microscopic details of the initial
conditions, in the case of a disturbance that connects two metastable
minima of a scalar potential, the large time evolution of an initial disturbance that crosses the
barrier is that of the critical bubble, in the limit of small $G_N$.  We expect
that the same is true for a potential that falls asymptotically to zero.
Once more, the underlying intuition is
that at late times, the energy released from the change of false to true vacuum
overwhelms any slight energy difference in the starting point.  So we expect the solution to go over to $\phi(\tau)$.  So in this section, we will focus principally on the behavior of the
critical bubble, $\phi(\tau)$.

\subsection{Field evolution with small $G_N$}

For small but finite $G_N$, there is a long period
where  $G_N \times T_{00} \times \tau^2 \ll 1$, gravitation is negligible, 
and the picture of the 
previous 
section of the flat-space evolution of the bubble (or disturbance) is unaffected.  For a vacuum bubble in de Sitter space, gravitational
effects become important, for fixed $r \ll H^{-1}$, for example, only once $t \sim H^{-1}$. Provided the bubble has evolved to a configuration approximately that of the
critical bubble, we can take over the critical bubble results (with gravity).

So we consider the bubble evolution in the region of Minkowski signature.  Writing the metric in the form
\beq
ds^2 = -d\tau^2 + \rho(\tau)^2 \left( d\sigma^2 + \sinh^2(\sigma) d\Omega_2^2 \right),
\eeq
the equations for $\rho$ and $\phi$ are:
\beq
\ddot \phi + 3 {\dot \rho \over \rho}\dot\phi + V^\prime(\phi) =0
\eeq
and
\beq
\dot \rho^2 = 1 + {\kappa \over 3} \left({1 \over 2}{\dot\phi}^2 +V(\phi) \right ) \rho^2.
\label{rhoequation}
\eeq

Note that if the bubble emerges in a region of large $\rho$ ($\kappa \rho^2 V \gg 1$) then, for the asymptotically falling potential, the kinetic term quickly comes to dominate in the equation for $\rho$; the system
becomes kinetic energy dominated.  This is visible in the numerical results we describe
subsequently.

We should pause here to consider the tunneling problem.  We will see in the next section that if we take the positive root in equation \ref{rhoequation}, one obtains an expanding universe in the future, but there is a singularity in the far past (before the appearance of the bubble).  Alternatively, if we take the negative root, the singularity appears in the far future.  Which root one is to take brings us to questions of the long-time history of the universe, i.e., how the universe came to be in the metastable false vacuum.  The point of our discussion in this paper is that this issue already arises classically. 

\subsection{Behavior of the equations for large $\tau$}

Before describing our numerical results, it is helpful to consider some crude approximations
which give insight into the behavior of the system.
In the region with $\xi = i \tau$, 
the equations become those of CDL in the time-like region:
\beq
\ddot \phi + 3 {\dot \rho \over \rho}\dot\phi + \frac{dV}{d\phi} =0,
\label{phiwithx1}
\eeq
\beq
\dot \rho = \pm \sqrt{ 1 +{\kappa \over 3} \rho^2 \left({1 \over 2} \dot \phi^2 + V(\phi)\right)}.
\label{phiwithx2}
\eeq
We argued at the end of the previous section
that we might expect that the potential is not particularly relevant in the $\phi$
equation for large $\rho(0)$.  Ignoring the potential, we can also ask, self consistently, whether the second term in the 
$\dot \rho$ equation dominates over the first.  If it does, we have an FRW universe with $k=0$ and 
\beq
\rho \propto (\tau-\tau_0)^{1/3},~~\tau>\tau_0;~~~~~~~~\rho \propto (\tau_0-\tau)^{1/3},~~\tau <\tau_0.
\eeq
(These are the results for a universe with $p=w\rho;~w=1$.)  We can see this directly from the equations.
We have
\beq
{\dot \rho \over \rho} = \pm\sqrt{\kappa \over 6} \dot \phi.
\eeq
So
\beq
\frac{d^2 \phi}{d\tau^2} \pm \sqrt{\frac{3\kappa}{2}} \dot \phi^2 = 0.
\eeq
We look for a solution of the form
\beq
\dot \phi = \alpha (\tau-\tau_0)^{-1},
\eeq
\beq
\alpha = \sqrt{\frac{2}{3\kappa}}.
\eeq
Plugging this back into the $\dot \rho$ equation gives
\beq
{\dot \rho \over \rho} = \pm {1 \over 3} {1 \over \tau - \tau_0},
\eeq
which is consistent with the expected $(\tau-\tau_0)^{1/3}$ behavior. 
So we have a singularity in the past or the future.

\begin{figure}[t!]
\includegraphics[width=0.55\linewidth]{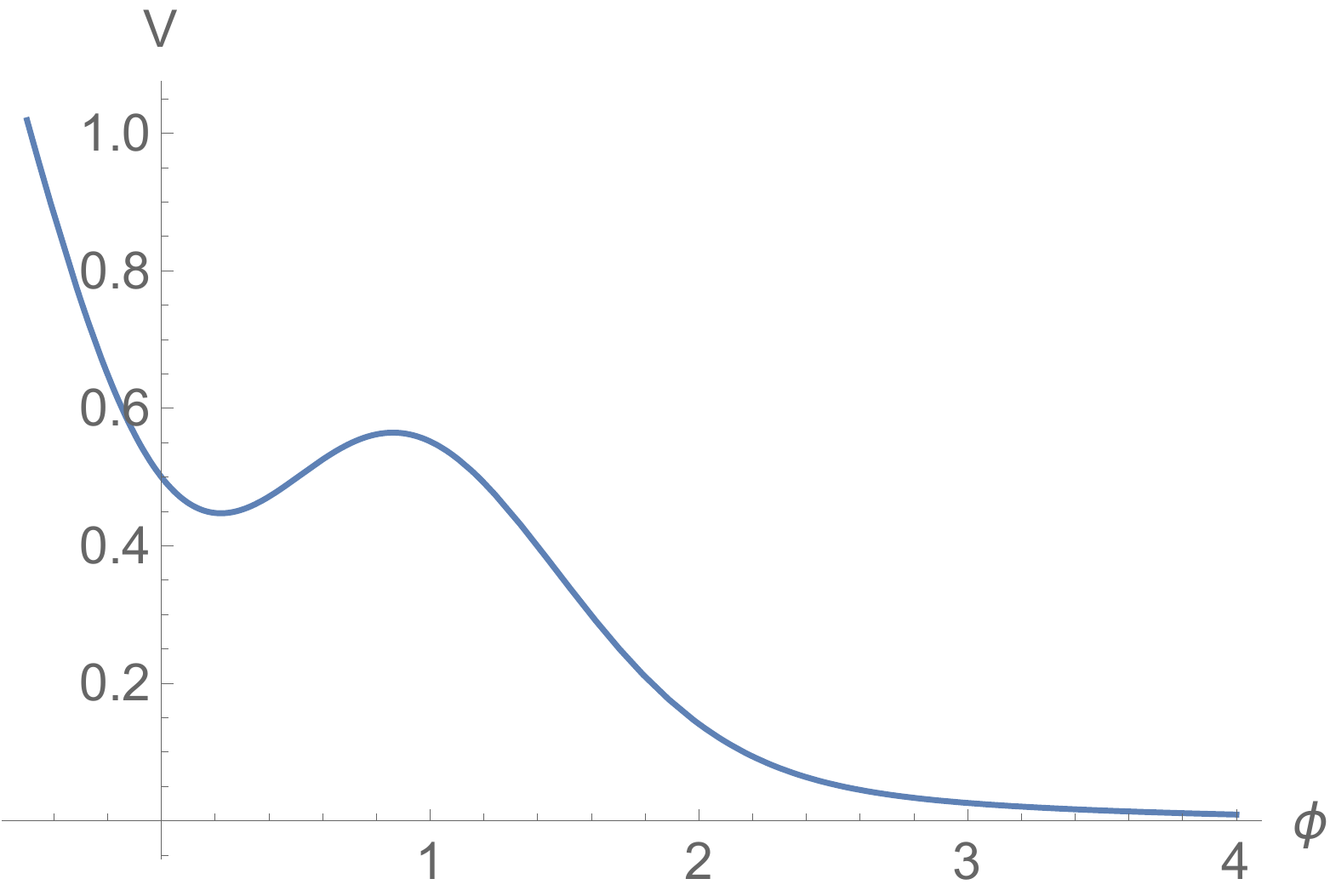}
\caption{
$\phi$ potential.
} 
\label{phipotential}
\end{figure} 

\begin{figure}[t!]
\includegraphics[width=0.55\linewidth]{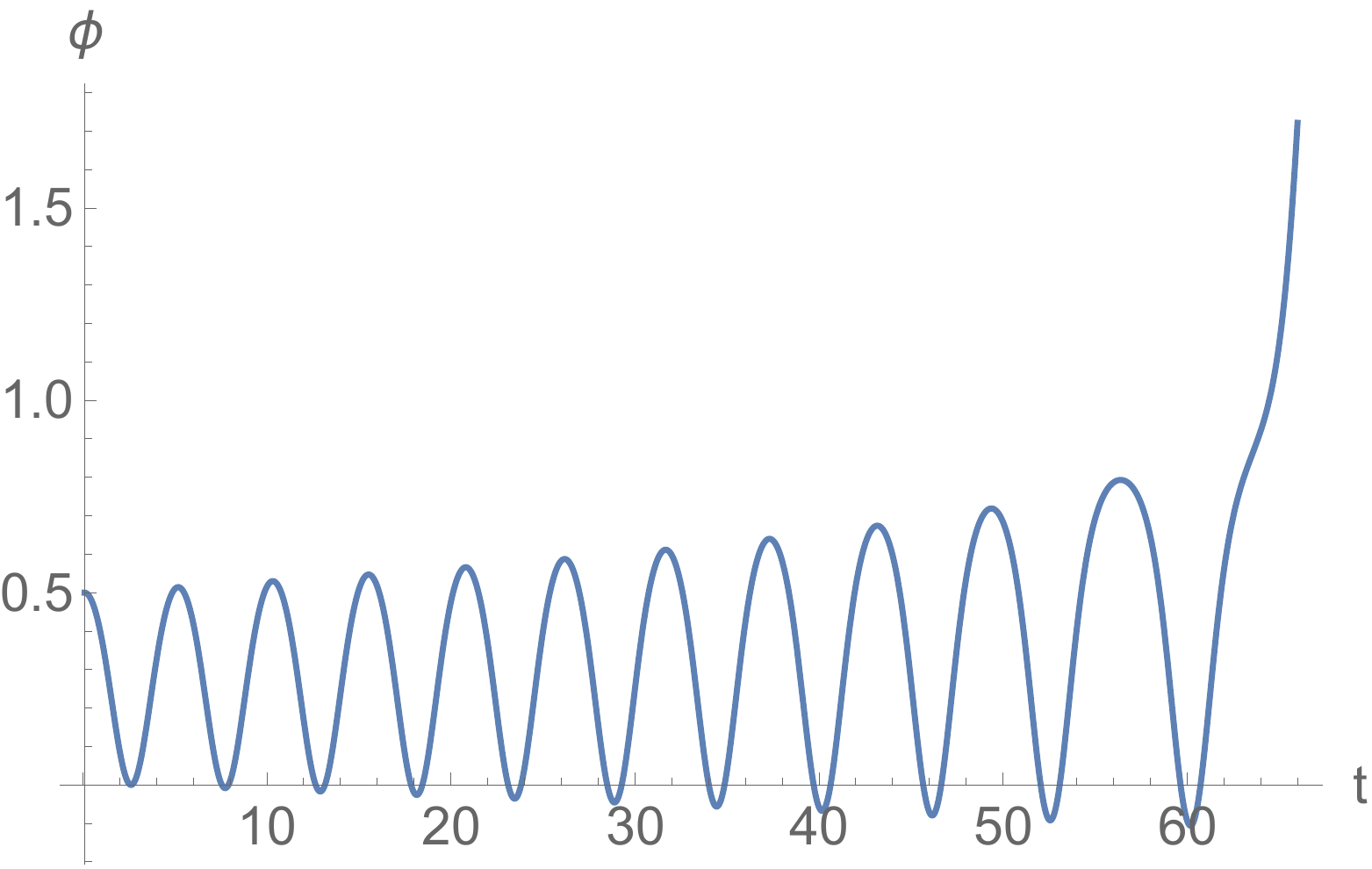}
\caption{
$\phi$ crosses the barrier.
} 
\label{phicrossing}
\end{figure} 

\begin{figure}[t!]
\includegraphics[width=0.55\linewidth]{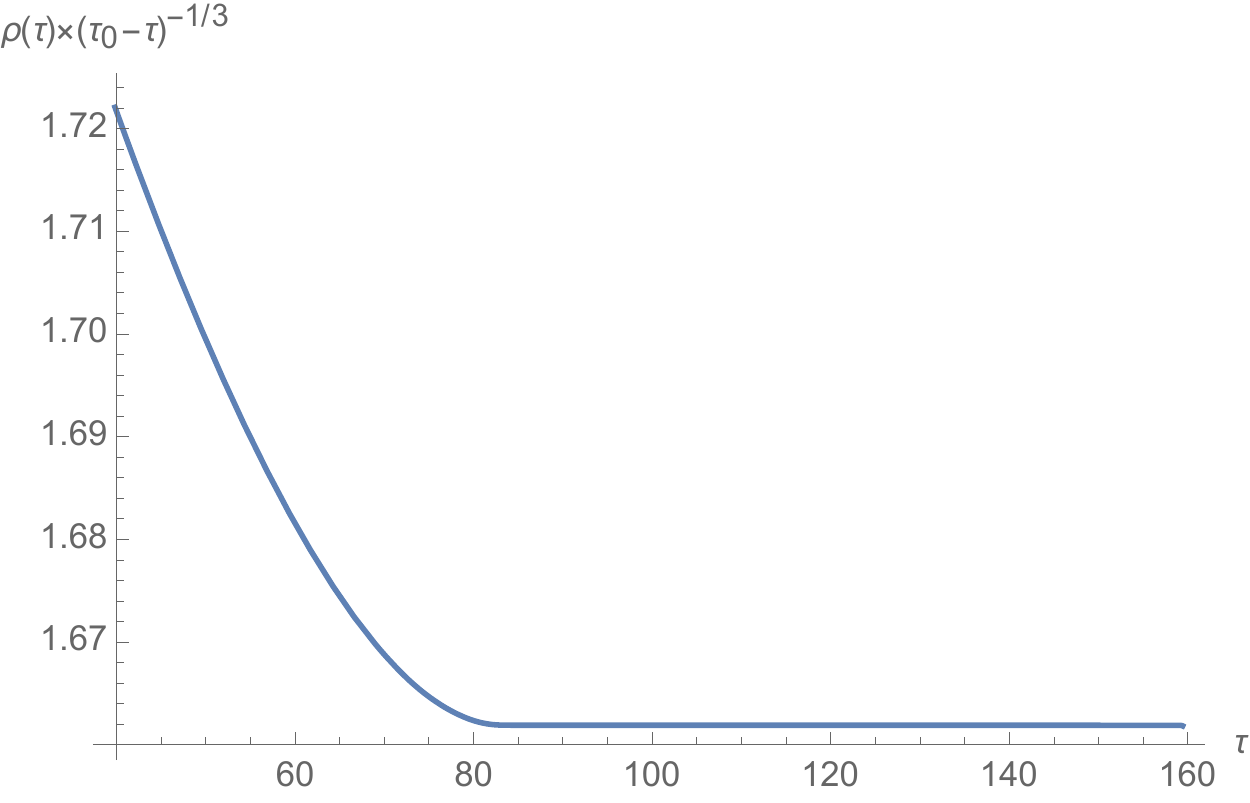}
\caption{
$\rho(\tau) \sim (\tau_0 - \tau)^{1/3};~\tau_0 \approx 159.5$.
}
\label{rhoweighted}
\end{figure}

For numerical studies, we designed a potential with a local de Sitter minimum that tends to zero for large $\phi$
\beq
V(\phi) = \frac{1}{2} e^{-\phi} + \phi^2 e^{-\phi^2},
\eeq

This is plotted in Figure \ref{phipotential}; the local minimum lies near $\phi=0.2$.
The potential blows up for negative $\phi$, but this will not concern us.
We solve equations \ref{phiwithx1} and \ref{phiwithx2} with $\phi_0$ taken to be
not too far from the local minimum, with small $d\phi / d\tau$ and with the negative sign in the root of the $\rho$ equation: $ \phi(\tau = 0) = 1/2; ~ \phi'(\tau = 0) = -10^{-6}; ~ \rho(\tau = 0) = 10$.  One sees (Figure \ref{phicrossing}) the scalar field roll over the barrier after some number of oscillations.  The ratio of potential to kinetic energy quickly tends to zero after the crossing.  As we expect, we find a singularity at a finite time in the future, and indeed $\rho(\tau)$ behaves as $( \tau_0-\tau)^{1/3}$ (Figure \ref{rhoweighted}).

We have argued that for more general initial conditions, provided gravity is sufficiently weak, the system evolves quickly to the bounce configuration with $G_N \approx 0$.  Its evolution will then be as above.

\subsection{Implications of the singularity}

Our main concern with the singularity is whether it is an obstruction to any sort of systematic analysis.
If we have a weak coupling, small curvature description of the system, allowing a perturbative
analysis, we expect to be able to write an effective Lagrangian including terms of successively higher dimension---higher numbers of derivatives---such as:
\beq
{\cal L} = \sqrt{g} \bigg( {1 \over G_N} {\cal R} + {\cal R}^2 + {1 \over M^2} {\cal R}^4 + \dots \nonumber
\\
+(\partial_\mu \phi)^2 + {1 \over M^4} (\partial_\mu \phi)^4  \bigg) .
\eeq
If one tries to analyze the resulting classical equations perturbatively, in the presence of $\dot \phi \sim 1 / (t-t_0)$ and ${\cal R} \sim 1 / (t-t_0)^2$, at low orders, the terms in the expansion
diverge and the expansion breaks down.  This is similar to the phenomena at a big bang or big
crunch singularity.

\section{Conclusions}

We have argued, from two points of view, that one cannot construct
de Sitter space in any controlled approximation in string theory.  First, we have seen that even allowing the possibility
of arbitrarily large fluxes, it is very difficult to find stationary points for which both the string coupling
is small and compactification radii are large, even before asking whether the corresponding cosmological
constant is positive or negative.  We have seen that typically when sensible stationary points exist, even if formally radii are large and couplings small, higher order terms in the expansions are not small.
Related observations have been made in \cite{oogurivafads}, based on conjectures about the behavior of quantum gravity systems. 

But our second obstacle seems even more difficult
to surmount:  a set of small perturbations of any would-be metastable de Sitter state, classically,
will evolve to uncontrollable singularities.

This is {\it not} an argument that metastable de Sitter states do not exist in quantum theories of
gravity; only that they are not accessible to controlled approximations.  The problem is similar
to the existence of big bang and big crunch singularities; we have empirical evidence that
the former exists in the quantum theory that describes our universe, but we do not currently have the tools
to describe these in a quantum theory of gravity.

Reference \cite{kachrukkltdesitter} has considered the question from the perspective of the KKLT \cite{kklt} constructions.  These involve vacua with fluxes, but the small parameter is not provided by taking all fluxes particularly large; 
rather, it arises from an argument that there are so many possible choices of fluxes that in some cases,
purely at random, there is a small superpotential.  In other words, there is conjectured to be a vast set of (classically) metastable states of which only a small fraction permit derivation of an approximate four-dimensional, weak coupling effective action.  Reference \cite{kachrukkltdesitter} argues that such a treatment is self consistent. 
We are sympathetic to the view that such an analysis provides evidence that if in some cosmology one lands for some interval in such a state, the state can persist for a long period.  But a complete description of such a cosmology is beyond our grasp at present.

In considering the cosmic landscape, one of the present authors has argued that, even allowing for the existence of such states in some sort of semiclassical
analysis, long-lived de Sitter vacua will be very rare, unless
protected by some degree of approximate supersymmetry \cite{dinestability}.  The breaking
of supersymmetry would almost certainly be non-perturbative in nature; searches for
concrete realizations of such states (as opposed to statistical arguments for the {\it existence} of such states, along the lines of KKLT) would be challenging.

Ultimately, at a quantum level, reliably establishing the existence of metastable de Sitter space 
appears to be a very challenging problem.
One needs a cosmic history, and it would be necessary that 
this history be under theoretical control, both in the past and in the future.  As a result, the significance of failing to find stationary points of an effective action describing metastable
de Sitter space is not clear.  We have seen that even thought of as classical configurations,
there are questions of stability and obstacles to understanding the system eternally, once small perturbations are considered. 
We view the question of the existence of metastable de Sitter space as an open one.

\vskip 1cm
\noindent
{\bf Acknowledgements:}  We thank Anthony Aguirre and Tom Banks for discussions.
This work was supported in part by the U.S. Department of Energy grant number DE-FG02-04ER41286.

\bibliography{desitter_space_in_string_theory}{}
\bibliographystyle{utphys}

\end{document}